\documentclass[aps,prx,preprint,bibliography,superscriptaddress]{revtex4-1}
\usepackage[utf8]{inputenc}
\usepackage{amsmath}
\usepackage{mathtools}
\usepackage{amsfonts}
\usepackage{amssymb}
\usepackage{braket}
\usepackage{graphicx}
\usepackage{multirow}
\usepackage{subfigure}
\usepackage{appendix}
\usepackage{comment}
\usepackage{textcomp}
\usepackage{color}
\usepackage[dvipsnames]{xcolor}
\usepackage{mathrsfs}
\usepackage{natbib}

\usepackage{lineno}

\def\lz{\mathcal{L}_0}
\def\lo{\mathcal{L}_1}

\begin{document}
\title{Characterizing Liouvillian Exceptional Points Through Newton Polygons and Tropical Geometry}
\author{P. Sayooj}
\email{sayoojp@iisc.ac.in}
\affiliation{Undergraduate Programme, Indian Institute of Science, Bangalore 560012, India}
\author{Awadhesh Narayan}
\email{awadhesh@iisc.ac.in}
\affiliation{Solid State and Structural Chemistry Unit, Indian Institute of Science, Bangalore 560012, India}


\date{\today}

\begin{abstract}
The dynamics of open quantum systems described by the Lindblad master equation follows according to non-Hermitian operators. As a result, such systems can host non-Hermitian degeneracies called Liouvillian exceptional points (EPs). In this work, we show that Newton polygons and tropical geometric approach allow identification and characterization of Liouvillian EPs. We use two models -- dissipative spin$-1/2$ system and dissipative superconducting qubit system -- to illustrate our method. We demonstrate that our approach captures the anisotropy and order of the Liouvillian EPs, while also revealing the subtle dependence on the form of the perturbation. Our analytical analysis is supplemented by direct numerical calculations of the scaling and exchange of eigenvalues around Liouvillian EPs. Our analytical approach could be useful in understanding and designing Liouvillian EPs of desired order.
\end{abstract}

\maketitle

\section{Introduction}

EPs are degeneracies unique to non-Hermitian systems~\cite{heiss2012physicsofEPs,bergholtz2021ExceptionalTopologyReview}. At EPs, both the eigenvalues and eigenvectors become degenerate, leading to intriguing properties. These include geometric phases, interesting phase transitions, and potentially enhanced sensitivity, among a plethora of other features~\cite{dembowski2004encircling,EPs_near_quantum_phasetrans,EPs_near_quantum_phasetrans,wiersig2016sensors,hodaei2017enhanced,chen2017exceptional,rosa2021exceptional,lai2019enhanced,banerjee2023non}. Notably, in the last few years, EPs have been controllably realized in a variety of platforms ranging from photonics , optics, and metamaterials~\cite{PTSymm_and_photonicEPs,EPsinOpticsandPhotonics,liu2017EPsinFanoGrapheneMetaMetarials}.

In addition to non-Hermitian Hamiltonians, EPs also emerge in open quantum systems described by the Lindblad master equation~\cite{minganti2019quantum,hatano2019exceptional,khandelwal2021LEP_inquantumthermalmachine,abo2024experimentalLEP,hashimoto2016LEPinLorentzgas,arkhipov2020LEPinlasertheory, seshadri2024liouvillian}. The effects of quantum jumps make these Liouvillian EPs distinct, yet related to the Hamiltonian EPs. There has been a growing interest in understanding the relation between the two types of EPs as well as the different properties of Liouvillian EPs. Recently, experimental signatures of Liouvillian EPs have been found in ultracold atomic platforms~\cite{ren2022chiral,sun2024encircling} and superconducting qubits~\cite{LEPSupercondQubits,chen2022decoherence,abo2024experimentalLEP}. Other platforms suggested to realize them include atomic vapors~\cite{kopciuch2025LEPinatomicvapour} and full counting statistics of open systems~\cite{pavlov2025topological}.

Recently, Newton polygons have been proposed to study Hamiltonian EPs~\cite{jaiswal2023characterizing}. Newton polygons were first formulated by Isaac Newton to obtain solutions of bivariable polynomial equations~\cite{Knorrerplanecurves}. They are now deeply embedded in algebraic geometry and various related fields in modern mathematics. In the context of non-Hermitian physics, the Newton polygon approach allows characterizing the EPs, along with associated non-Hermitian phenomena, such as the non-Hermitian skin effect~\cite{jaiswal2023characterizing}.

Tropical geometry is a relatively modern branch of mathematics lying at the interface between algebraic geometry and polyhedral geometry~\cite{maclagan2015introtoTropicalGeometry,mikhalkin2006tropicalGeo}. Broadly, it aims to study solutions of polynomial equations by using a tropical arithmetic. Furthermore, tropical ideas have also found interesting applications in a wide variety of fields including economics, genetics, and physics~\cite{kapranov2011Tropgeointhermodynamics,samal2015MathBio,noel2012tropgeoinBioChem,gross2011TropInMirrorSymm}. Very recently, a unified tropical geometry framework has been proposed to study Hamiltonian EPs and their properties~\cite{banerjee2023tropical}.

In this paper, we show that Newton polygons and tropical geometric approach can directly characterize Liouvillian EPs. We use two models -- dissipative spin$-1/2$ system and dissipative superconducting qubit system -- to elucidate our method. Using the dissipative spin$-1/2$ model we show that our method characterizes the intrinsic EP property of the inequality between the geometric and algebraic multiplicity and that the amoeba describes the scaling property of the Liouvillian EPs. Furthermore, using the dissipative superconducting qubit model we show that our approach captures the anisotropy and order of the Liouvillian EPs depending on the form of the perturbation. We supplement our Newton polygon and amoeba analyses with calculations of scaling of the eigenvalues as well as their exchange upon circling around the Liouvillian EPs. We are hopeful that our analytical approach will be useful in furthering our understanding of Liouvillian EPs from both theoretical and experimental directions.

The rest of this paper is organized as follows: We briefly introduce the mathematical basis behind our method in Sec.~\ref{sec:recap}. We shall consider two different models of Liouvillian systems where exceptional topology is observed. These are detailed in Sec.~\ref{sec:model1} (dissipative spin$-1/2$ system) and Sec.~\ref{sec:model2} (dissipative superconducting qubit system). Finally, we summarize our work in Sec.~\ref{sec:summary}.

\section{Recap of Newton Polygons and Tropical Algebra}\label{sec:recap}

We begin by briefly introducing the method of Newton polygons. Using this as a basis, we will then motivate towards the concepts of tropical geometry and an alternative way to obtain the order of a Liouvillian EP using a construction known as the amoeba of the algebraic variety.

\subsection{Newton-Puiseux Theorem and the Newton Polygon Method}

The Newton-Puiseux theorem states that the field of Puiseux series is algebraically closed~\cite{NewtonPuiseuxThm}. Recall that a Puiseux series in $\epsilon$ is an infinite sum of the form

\begin{equation}
\sum_{k=k_0}^\infty c_k \epsilon^{\frac{k}{n}},
\end{equation}

with fixed $n$ and where $c_i$ belongs to the concerned field, which, here, is the field of complex numbers. The lowest power $\frac{k_0}{n}$ is called the valuation of the Puiseux series. We will revisit this term subsequently in the subsection on tropical geometry in Section II B. From the Newton-Puiseux theorem it follows that for any algebraic polynomial in two variables of the form $f(\epsilon,\omega)$, all solutions for $\omega$ can be written as a Puiseux series in $\epsilon$. This is because such an algebraic polynomial can be written as a polynomial in a single variable $\omega$ over the fields of Puiseux series in $\epsilon$. This field, being an algebraically closed one, by definition implies that all solutions of the polynomial belong to the field itself.

Due to Issac Newton, Newton polygons are an effective method to solve for these Puiseux series~\cite{howtocomputepuiseux} . We are however only concerned with the so-called valuation of the solutions as these are the orders of the EPs, as we will soon see. For a given polynomial $f(\omega,\epsilon)$, one writes this as a polynomial in $\omega$ in the field of Puiseux series in $\epsilon$, that is $f(\omega,\epsilon)= \sum_{i} c_i(\epsilon)\omega^{i}$, where $c_i(\epsilon)$ is a Puiseux series in $\epsilon$. This we can do because a polynomial is also a finite Puisuex series. For each non-zero $c_i$, we plot $(i,\text{val}(c_i))$, where $\text{val}(c))$ refers to the valuation of $c$ defined in the previous paragraph, and obtain its convex hull. This region is called the Newton polygon of $f$. For simple algebraic equations, this is equivalent to plotting $(i,j)$ for every non-zero term of the form $(\omega^{i},\epsilon^{j})$ and taking its convex hull. Indeed this is what we will be doing for our characterization of Liouvillian EPs.

By segments of the polygon, we will refer to those edges that are to the left and below every point inside the polygon. It is a central result that every solution of $f$ in $\omega$ can be written as a Puiseux series in $\epsilon$ such that its valuation (lowest power of $\epsilon$) will be the negative of one of the slopes of the segments. Furthermore, the number of solutions with this lowest power is equal to the projection of the segment on the horizontal axis.

Let us next briefly revisit the key ideas behind Liouvillian operators to set the stage for connecting to Newton-Puiseux expansions. A Liouvillian operator usually acts on the matrix space of the density operators~\cite{manzano2020IntroToLindblad}. 

However, by flattening an $n \times n$ density operator into a column vector of $n^2$ components, one obtains the Liouvillian operator in the form of a $n^2 \times n^2$ matrix $\mathcal{L}_0$. A perturbation can also be represented in the same way as proportional to a matrix $\mathcal{L}_1$.

An exceptional point occurs when the algebraic multiplicity of a particular eigenvalue exceeds the geometric multiplicity \cite{a7referee}. The best accepted definition of an $m$-th order EP is when $m$ \textit{eigenvectors} coalesce into a single one. Note that we use eigenvectors to denote the lack of geometric multiplicity. This definition is equivalent to having a Jordan block of size $m$ in the Jordan decomposition of the matrix, which in our case is the Liouvillian.

As mentioned before, under a generic perturbation $ \epsilon \mathcal{L}_1$, the associated eigenvalues of an EP-$m$ diverge as $\omega-\omega_0 \sim \epsilon^{1/m}$ \cite{kato2013perturbation,a3referee}. Thus, the solutions of the algebraic equation
\begin{equation}
    \det(\mathcal{L}_0 +\epsilon \mathcal{L}_1- I(\omega-\omega_0)) = 0,
    \label{char}
\end{equation}

in $\omega$ will be a Puisuex series of $\epsilon$ with the valuation $\frac{1}{m}$ and can be easily detected from the Newton polygon. We emphasize that when the eigenvalues diverge in the reciprocal order of its EP order, the perturbation is said to be a generic perturbation. Such perturbation matrices satisfy no fine-tuned relations. From our method, one can directly see that a \emph{non-generic} perturbation matrix is one that causes the vanishing of certain terms in the characteristic equation to actively change the structure of the Newton polygon. To change the structure of the Newton Polygon, certain coefficients of the characteristic polynomials should be set to zero. These coefficients will be some complex polynomials in the matrix entries of $\mathcal{L}_1$. If these coefficients are given by $g_1(\mathcal{L}_1),g_2(\mathcal{L}_1) \cdots g_k (\mathcal{L}_1)$ then the set of all non-generic matrices will be given by the union of their roots. Hence, in a measure theoretic sense, most perturbation matrices are generic, and so the easiest way to generate one is to draw a random $N \times N$ matrix, where $N=n^2$, with $n$ being the dimension of the concerned Hilbert space.

This anisotropy of EPs with respect to the perturbation is further verified by studies of deformations~\cite{a4referee} and Hessenberg classification of $\mathcal{L}_1$~\cite{a1referee,ma1998nongeneric}. The anisotropy of Hamiltonian EPs is well documented in non-Hermitian Hamiltonian systems~\cite{grom2025graph,ding2018experimentalanisotropy,demange2012signatures}. This is described in detail in the Appendix in connection with the Jordan decomposition. 

\subsection{Tropical Geometry and amoebas}

Next, we briefly recapitulate the central ideas of tropical algebra. Tropical geometry is based on the idea of a tropical semiring~\cite{maclagan2015introtoTropicalGeometry}. This is defined usually in the minimum ("min") convention. The min tropical semiring $\mathbb{T}$ is the semiring \((\mathbb{R} \cup \{+\infty\}, \oplus, \otimes)\), with the operations

\begin{equation}
x \oplus y = \min\{x,y\}, \quad 
x \otimes y = x + y.
\end{equation}

We then define a valuation of a field $\mathbb{K}$, as a function

\begin{equation}
\operatorname{val} : \mathbb{K} \to \mathbb{R} \cup \{\infty\},    
\end{equation}

such that the following properties hold

\begin{itemize}
    \item $\operatorname{val}(a) = \infty$ if and only if $a = 0$;
    \item $\operatorname{val}(ab) = \operatorname{val}(a) + \operatorname{val}(b)$;
    \item $\operatorname{val}(a+b) \geq \min\{\operatorname{val}(a), \operatorname{val}(b)\}$ for all $a,b \in \mathbb{K}$.
\end{itemize}

In our framework, we are only concerned with the field of Puiseux series with coefficients in the complex numbers $\mathbb{C}$. This field has a natural well-defined valuation which is given by taking a non-zero Puiseux series to the lowest exponent that appears in its expansion. This is indeed what we will be calling the valuation of the solution in the rest of the article. 

In its most basic form, tropical geometry gives a method to compute the valuations of the non-zero roots of a non-zero polynomial $p \in \mathbb{K}[\lambda]$ in terms of the valuations of the coefficients of $p$. More precisely, given a non-zero polynomial \(p = \sum_{i=0}^d a_i \lambda^i \in \mathbb{K}[\lambda]\), $a_i \neq 0$, its tropicalization \(\operatorname{trop}(p) : \mathbb{R} \to \mathbb{R}\) is defined as

\begin{equation}
\operatorname{trop}(p)(\omega) = \min_i \{ \operatorname{val}(a_i) + i \cdot \omega \}.
\end{equation}

We will use this idea to characterize the Liouvillian EPs within the tropical framework.

An important construction in tropical geometry is the amoeba of an algebraic variety due to Gelfand, Kapranov and Zelevinsky~\cite{AmoebaFirstPaper}. A complex algebraic variety of a set of polynomials in $n$ variables is the set of all complex valued solutions $V \subseteq (\mathbb{C}^\star)^n = (z_1,\ldots,z_n)$, all of whose coordinates are non-zero.

Let \(\operatorname{Log} : (\mathbb{C}^\star)^n \to \mathbb{R}^n\) be the logarithmic map that takes $(z_1,\ldots,z_n)$ to $(\log(|z_1|), \ldots, \log(|z_n|))$. The amoeba of $V$ is the image of the logarithmic map restricted to $V$. A related and important concept is the spine of the amoeba, and is defined as the limit as $t \to \infty$ of the parameterized logarithmic map

\begin{equation}
\operatorname{Log}_t(z_1, \ldots, z_n) = (\log_t(|z_1|), \ldots, \log_t(|z_n|)).
\end{equation}

\begin{figure}[h!]
    \centering
    \includegraphics[width=0.95\linewidth]{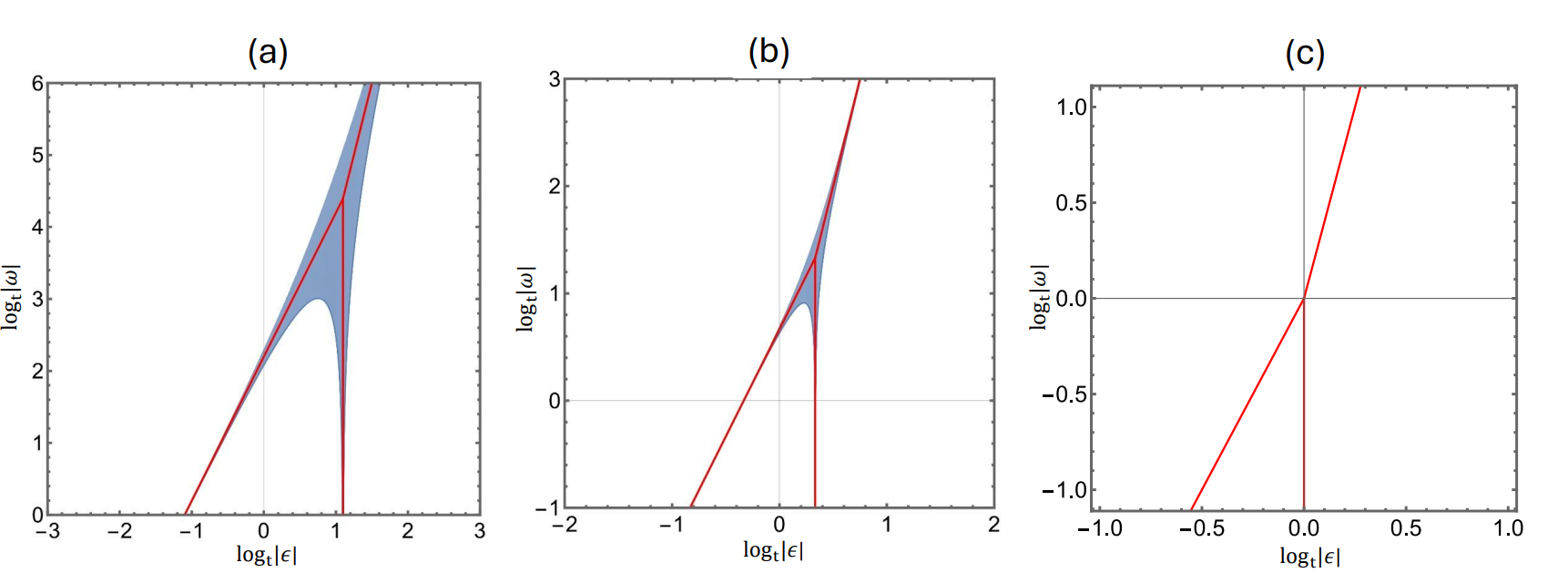}
    \caption{\textbf{Illustration of amoebas and their spine.} Plot of the logarithmic function on the set of solutions of the polynomial equation $f(\omega,\epsilon)= -\epsilon + 9\omega^2 -  \omega^4$. This is effectively the polynomial in Model I. The base of the logarithm goes as $t=e,\, t= 10 e, \,t=\infty$ in (a), (b), and (c), respectively. The tentacles of the amoeba are visible as they branch out to infinity. The asymptotic lines of the tentacles are drawn in red and they have the equations: $y=2x + \log_t{9},\: y=4x,$ and $x=\log_t{3}$. As $t\rightarrow \infty$ only the asymptotes remain and they coincide with the spine of the amoeba. Detailed analysis of obtaining this amoeba is given in the Appendix.}
    \label{fig:amoebaspine}
\end{figure}

The amoeba of a typical polynomial contains unbounded rays that are called its tentacles like in Figure \ref{fig:amoebaspine}. These tentacles asymptotically coincide with the spine. A central result concerning them is that the slope of these tentacles is the set of normal to the outer edges of the Newton polygon~\cite{viro2002amoeba}. We will be thus applying this method to the characteristic polynomial associated with the Liouvillian. Since amoebas can be extended to higher dimensions, these could potentially also be used to analyze EPs in higher-dimensional parameter spaces ~\cite{a5referee} or under non-linear conditions.

\section{Liouvillian Exceptional Points in Spin$-1/2$ Dissipative System}\label{sec:model1}

Let us first consider a simple, albeit quite general, model of a spin$-\frac{1}{2}$ particle, with the Hamiltonian

\begin{equation}
    \hat{H}=\frac{\Omega}{2} \hat{\sigma_z},
\end{equation}

with $\Omega$ denoting the energy splitting between the levels. We consider the case in which the system evolves under the action of three decay channels, $(\hat{\sigma}_x, \hat{\sigma}_y, \text{ and } \hat{\sigma}_-)$. Here $\gamma_x,\gamma_y$ and $\gamma_-$ denote the dissipation rates in these channels, respectively. This system is described by the Lindblad master equation and was recently studied by Minganti \textit{et al.} in this context of Liouvillian EPs~\cite{minganti2019quantum}. The corresponding Lindblad master equation is written as

\begin{equation}
\mathcal{L} \hat{\rho}(t) = -i [\hat{H}, \hat{\rho}(t)] + \gamma_-\mathcal{D}[\hat{\sigma}_-] \hat{\rho}(t) + \gamma_x \mathcal{D}[\hat{\sigma}_x]\hat{\rho}(t)+ \gamma_y \mathcal{D}[\hat{\sigma}_y]\hat{\rho}(t),
\end{equation}

where $\hat{\sigma}_{x,y,z}$ are the Pauli matrices, $\hat{\sigma}_{\pm} = (\sigma_x \pm i \sigma_y)/2$ and $\mathcal{D}$ is the usual Lindblad dissipator superoperator given by $ \mathcal{D}[\hat{\Gamma}] = \hat{\Gamma} \cdot \hat{\Gamma}^\dagger 
- \tfrac{1}{2} \{ \hat{\Gamma}^\dagger \hat{\Gamma}, \, \cdot \}$. Now, by flattening the density matrix,

\begin{equation}
\left(
\begin{array}{cc}
 \rho _{1,1} & \rho _{1,2} \\
 \rho _{2,1} & \rho _{2,2} \\
\end{array}
\right)\to \left(
\begin{array}{c}
 \rho _{1,1} \\
 \rho _{1,2} \\
 \rho _{2,1} \\
 \rho _{2,2} \\
\end{array}
\right),
\end{equation}

one obtains the $4 \times 4$ Lindbladian superoperator matrix as

\begin{equation}
\mathcal{L}_0=
\left(
\begin{array}{cccc}
 \gamma _-+\gamma _x+\gamma _y & 0 & 0 & \gamma _x+\gamma _y \\
 0 & -\frac{1}{2}\gamma _-+\gamma _x+\gamma _y-i \omega  & \gamma _x+\gamma _y & 0 \\
 0 & \gamma _x+\gamma _y & -\frac{1}{2}\gamma _-+\gamma _x+\gamma _y+i \omega  & 0 \\
 \gamma _-+\gamma _x+\gamma _y & 0 & 0 & \gamma _x+\gamma _y \\
\end{array}
\right).
\end{equation}

We will next demonstrate our method, which allows us to obtain the exhaustive set of all Liouvillian EPs.

\subsection{Newton polygon and tropical analysis}

Using Newton polygons, we are able to search for all Liouvillian EPs. We consider the characteristic polynomial 

\begin{equation}
\det(\mathcal{L}_0+\epsilon\mathcal{L}_1-I(\omega+\omega_0)) = 0,
\end{equation}

where $\lo$ denotes a perturbation to the system Lindbladian. We consider generic perturbations of strength $\epsilon$, that is, cases when the eigenvalues scale as $\epsilon^{\frac{1}{n}}$ for $n$ coalescing eigenvectors. In particular $\mathcal{L}_1$ can be any generic perturbation. An efficient way to do this is to use a randomly generated perturbation matrix. Here $I\omega_0$ is a shifting term to move the Liouvillian EPs to zero energy for simplicity. 

Now to scan for second-order Liouvillian EPs, we equate the coefficient of $\omega$ and the constant term to be zero. The detailed expressions for these coefficients are provided in the Appendix~\ref{sec:appendix}.

Note that $\omega_0$ is a variable that we have introduced, which can be eliminated using the resultant. We remind that the resultant of two polynomials is the determinant of their Sylvester matrix~\cite{ResultantBook}. It evaluates to zero if and only if the two polynomials have a common root. This way, $\omega_0$ can be eliminated and one can obtain the exhaustive set of all points where two eigenvalues are equal. One can now solve for say $\gamma_x$ to obtain different parameter regimes for the Liouvillian EPs. Some of these turn out to be diabolic points, that is, the familiar Hermitian degeneracies with algebraic multiplicity equal to the geometric multiplicity~\cite{berry1984diabolical}. It is well known that for such points, the eigenvalues and eigenvectors disperse linearly under a perturbation.

To filter them out we obtain the Newton polygon and check that the slope is more than $-1$. Hence, we find three sets of Liouvillian EPs for the dissipative spin$-1/2$ system

\begin{eqnarray}
\gamma_x=\gamma_y-\Omega\quad \text{at} \quad \omega_0=-\frac{1}{2}\gamma_--2\gamma_y+\Omega, \\
\gamma_x=\gamma_y+\Omega\quad \text{at} \quad\omega_0=-\frac{1}{2}\gamma_--2\gamma_y-\Omega, \\
\gamma_x=-\frac{1}{2}\gamma_--\gamma_y \quad \text{at} \quad \omega_0=0.
\end{eqnarray}

We note that conventionally the dissipation rates are assumed to be non-negative, hence the last region, while mathematically allowed, is not directly physical. The dissipation rates represent the irreversible loss of energy and information in Lindbladian systems, and as such they should always be positive. Nevertheless, we note that our method finds the exhaustive set of Liouvillian EPs within the chosen parameter space.

\begin{figure}[!h]
    \fbox{\includegraphics[width=0.95\textwidth]{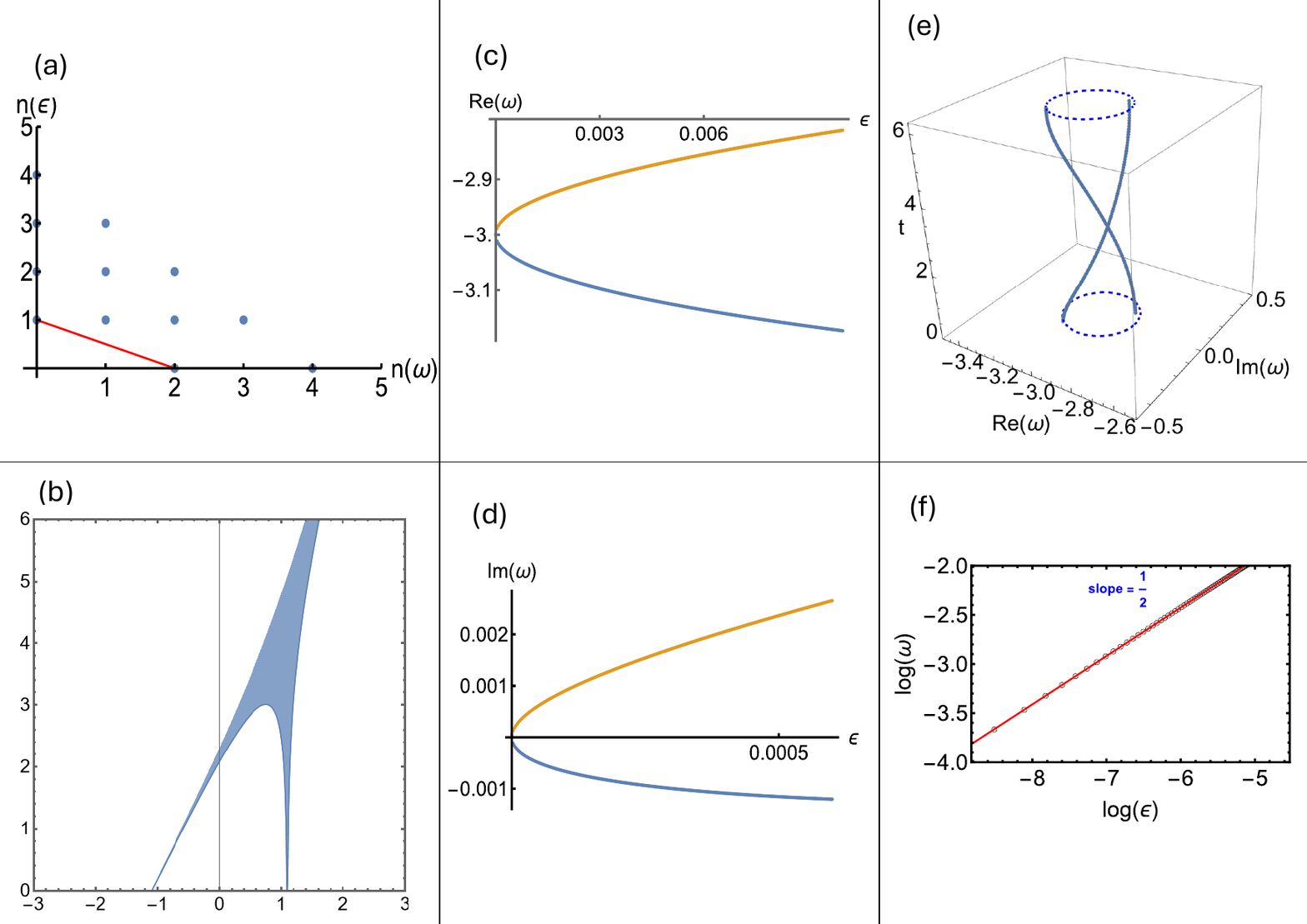} }
    \caption{\textbf{Second-order Liouvillian exceptional point in dissipative spin$-1/2$ model.} (a) The Newton polygon corresponding to the Liouvillian of the dissipative spin$-1/2$ system. It shows a slope of $-\frac{1}{2}$, thus indicating a second-order Liouvillian EP. (b) The amoeba of the characteristic polynomial at the Liouvillian EP. A tentacle branches out with a slope value of 2, which matches the prediction by Newton polygons and thus further indicates a square root topology. The (c) real and (d) imaginary parts of eigenvalues near the Liouvillian EP as a function of the perturbing parameter $\epsilon$. (e) The encircling of the Liouvillian EP reveals the characteristic exchange of the eigenvalues as a function of the parameter $t$, where we used $\lz+0.01\lo e^{it}$. (f) Plot of $\log(\omega-\omega_0)$ with respect to $\log (\epsilon)$, where the data points are taken from (c) and (d), illustrating the square root behaviour near the Liouvillian EP. The other parameters are chosen to be $(\gamma_-=0,\gamma_x=1,\gamma_y=2, \Omega=1)$.} 
    \label{fig:spin half model}
\end{figure}

Let us take one of the points of first set. The features of this Liouvillian EP are presented in Figure~\ref{fig:spin half model}. The corresponding Newton polygon and amoeba are shown in Figure~\ref{fig:spin half model} (a) and (b), respectively. We find that the amoeba shows a tentacle with a slope of 2, thus signifying a second-order Liouvillian EP. This is consistent with our findings using our Newton polygon approach. We have further confirmed this for the Liouvillian EPs in the other sets as well.

To further characterize the Liouvillian EPs and verify our Newton polygon and tropical approaches, we have carried out direct diagonalization of the Liouvillian to obtain the real and imaginary parts of the eigenvalues near the Liouvillian EP, which are shown in Figure~\ref{fig:spin half model}(c) and (d), respectively. The numerical analysis on the perturbation of eigenvalues was done by direct diagonalization using Eigenvalues function implemented in Mathematica. We find that the eigenvalues indeed disperse in a square root fashion as a function of the perturbing parameter $\epsilon$ (Figure~\ref{fig:spin half model}(f)). As an additional check, we encircle the Liouvillian EP by tuning a parameter $t$, such that $\lz+0.01\lo e^{it}$ with $t\in [0,2\pi]$. As shown in Figure~\ref{fig:spin half model}(e), we observe that the eigenvalues swap between themselves as the Liouvillian EP is encircled. This follows as a consequence of the $\epsilon^{1/N}$ scaling, where $N$ eigenvalues undergo cyclic perturbations when an $N$-th order EP is circled. To summarize, we find that using our combined approach of Newton polygons and tropical geometry we are able to fully characterize the set of Liouvillian EPs in a fairly general dissipative model.

\section{Liouvillian Exceptional Points in Dissipative Superconducting Qubit System}\label{sec:model2}
\def\lj{\mathcal{L}_{\text{jumps}}}

To further illustrate our approach, as the next example, we study a dissipative qubit system. We motivate the model based on the recent experiments by Chen \textit{et al.}, who studied the dynamics of a superconducting qubit, which was perturbed by quantum jumps between the energy levels~\cite{LEPSupercondQubits}. In addition to observing a third-order EP, they also revealed an enhancement of decoherence arising from quantum jumps in the vicinity of the EP~\cite{LEPSupercondQubits}.

In the experiments of Chen \textit{et al.}, the qubit was obtained by considering the two higher energy states of a superconducting qutrit system with the states $\ket{e}, \ket{f}$ and $ \ket{g}$ arranged in decreasing order of energy~\cite{LEPSupercondQubits}. A microwave drive with an amplitude $J$ was applied between the $\ket{e}$ and $\ket{f}$ states. There are two possible channels of decay -- spontaneous emission from $\ket{e}$ to $\ket{f}$ and from $\ket{f}$ to $\ket{g}$ -- with rates $\gamma_e$ and $\gamma_f$, respectively. Hence, the Lindblad master equation for the three-state system is given by

\begin{equation}
\frac{\partial \rho_{\mathrm{tot}}}{\partial t}
= -i [H_c, \rho_{\mathrm{tot}}]
  + \sum_{k = e,f} \left(
  L_k \rho_{\mathrm{tot}} L_k^{\dagger}
  - \frac{1}{2} \{ L_k^{\dagger} L_k, \rho_{\mathrm{tot}} \}
  \right),
\tag{4}
\end{equation}

where $\rho_{\mathrm{tot}}$ denotes a $3 \times 3$ density operator. The Lindblad dissipators, $L_e = \sqrt{\gamma_e}\,|g\rangle\langle e|$ and
$L_f = \sqrt{\gamma_f}\,|e\rangle\langle f|$, describe the decay from
$|e\rangle$ to $|g\rangle$ and from $|f\rangle$ to $|e\rangle$, respectively.

Further, a microwave drive with amplitude $J$ was applied to the $\{|e\rangle, |f\rangle\}$ submanifold, and in the rotating frame, one obtains~\cite{LEPSupercondQubits}

\begin{equation}
H_c = J (|e\rangle\langle f| + |f\rangle\langle e|).
\end{equation}

The dynamics of the qubit system, therefore, will be given by a hybrid Liouvillian superoperator as

\begin{equation}
\frac{\partial \rho}{\partial t} = \left( \mathcal{L}_0 + \mathcal{L}_{\text{jumps}} \right)\rho =\mathcal{L}_{\text{eff}} \rho.
\end{equation}

As before, we perform the flattening procedure to convert this operator defined on the space of $2\times2$ matrices into a four-dimensional vector space to obtain,

\begin{equation}
\lz=
\begin{pmatrix}
-\gamma_{e} & iJ & -iJ & 0 \\
iJ & -(\gamma_{e} + \gamma_{f})/2 & 0 & -iJ \\
- iJ & 0 & -(\gamma_{e} + \gamma_{f})/2 & iJ \\
0 & -iJ & iJ & -\gamma_{f}
\end{pmatrix},
\end{equation}

\begin{equation}
\lj=\begin{pmatrix}
0 & 0 & 0 & \gamma_{f} \\
0 & 0 & 0 & 0 \\
0 & 0 & 0 & 0 \\
0 & 0 & 0 & 0
\end{pmatrix}.
\end{equation}

\subsection{Newton polygon and tropical analysis}

\begin{figure}[!h]
   \fbox{\includegraphics[width=0.95\textwidth]{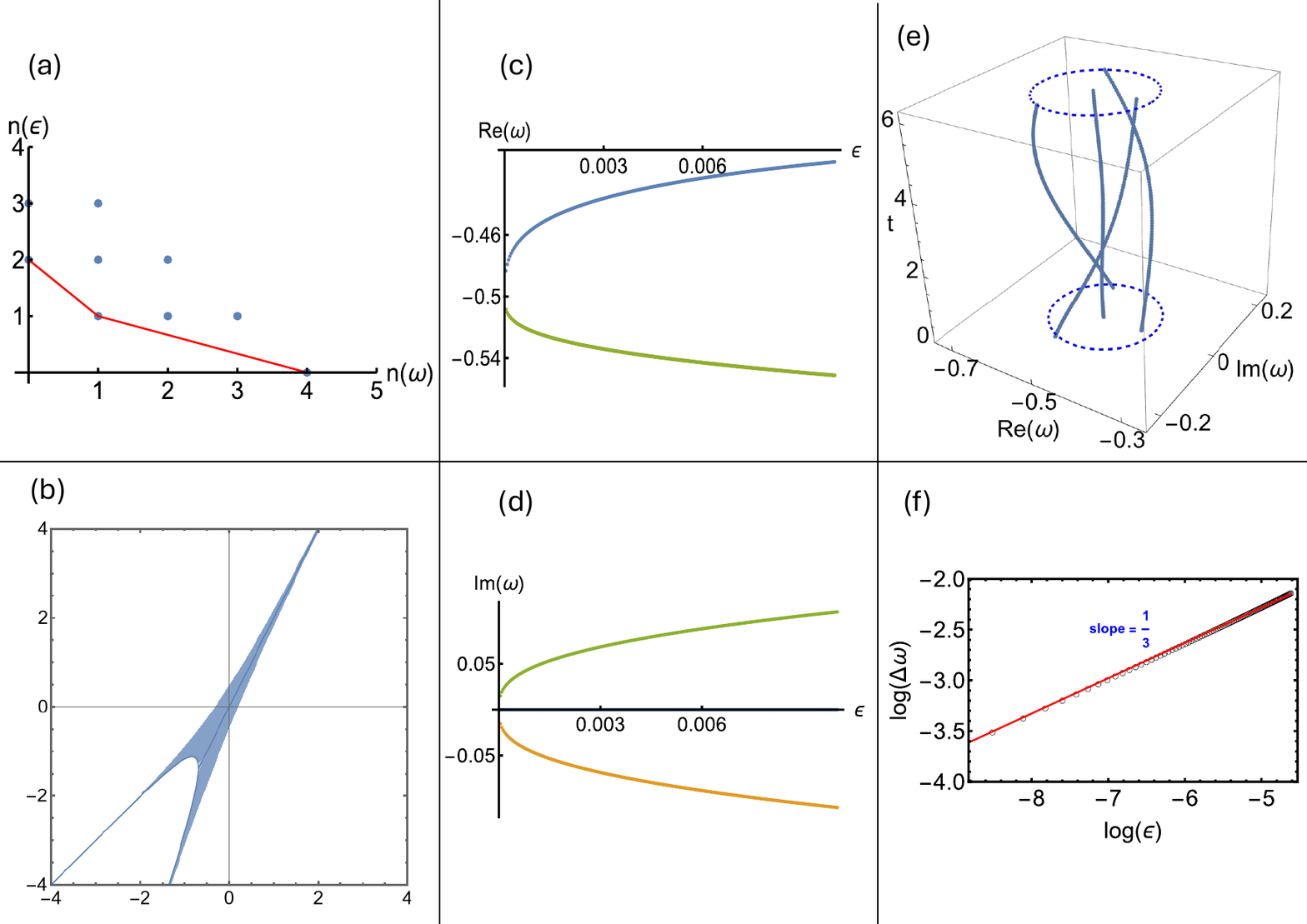} }
    \caption{\textbf{Hybrid Liouvillian exceptional points in dissipative qubit system with $\gamma_f$ perturbation.} (a) The Newton polygon exhibiting two segments with non-trivial slopes $-1$ and $-\frac{1}{3}$, respectively. (b) The amoeba of the shifted characteristic polynomial at the Liouvillian EP. We note two tentacles branching out to negative infinity below with slopes 1 and 3, precisely normal to the segments of the Newton polygon. The (c) real and (d) imaginary parts of the numerically evaluated energy eigenvalues near the Liouvillian EP showing three eigenvalues (green, orange, and blue) coalescing. As seen in panel (c) and (d), two of the eigenvalues (green and orange) are complex conjugates of each other, while the third eigenvalue (blue) is purely real. (e) The encircling of the Liouvillian EP, i.e., a plot of eigenvalues of the matrix $\mathcal{L}_{eff}+0.01\lo e^{it}$ with varying $t$. (f) Plot of $\log(\omega-\omega_0)$ vs $\log (\epsilon)$, where the data points are taken from the orange curve in (c) and (d). A slope of $\frac{1}{3}$ indicates a cube root scaling. The model parameters are chosen to be ($\gamma_f=0,\gamma_e=1,J=0.25$), and the perturbation matrix is $\frac{\partial}{\partial \gamma_f}\mathcal{L}_{eff}$ to consider small variations in $\gamma_f$. Note that $\mathcal{L}_{eff}$ is linear in $\gamma_f$.} \label{fig: Gamma F perturb}
\end{figure}

Let us next use this model of dissipative qubits to demonstrate the anisotropy shown by Liouvillian EPs. We first consider the case with $\gamma_f=0$. By an analogous procedure followed in the previous section, we find a Liouvillian EP at the point given by ($\gamma_f=0,\gamma_e=1,J=0.25$) at $\omega_0=-\frac{1}{2}$. At this point the Liouvillian superoperator matrix takes an interesting form as can be see from its Newton polygon (Figure~\ref{fig: Gamma F perturb}). The Jordan canonical form of the superoperator is 

\begin{equation}
\begin{pmatrix}
-\tfrac{1}{2} & 0 & 0 & 0 \\
0 & -\tfrac{1}{2} & 1 & 0 \\
0 & 0 & -\tfrac{1}{2} & 1 \\
0 & 0 & 0 & -\tfrac{1}{2}
\end{pmatrix}.
\end{equation}

Here, we have a fourth-order degeneracy of the eigenvalues in the characteristic equation, but instead of only one, we obtain two different linearly independent eigenvectors. The Jordan canonical form shows two blocks with size $1\times1$ and $3 \times3$. Such EPs were observed before in two-dimensional plaquette systems~\cite{a6referee}. This means that under a generic perturbation, the eigenvalues disperse in these groups. We first consider perturbations due to quantum jumps, that is, when the value of $\gamma_f$ is perturbed. The corresponding matrix $\lo$ is obtained as $\lo=\frac{\partial}{\partial\gamma_f} \mathcal{L}_{\text{eff}}$. Figure~\ref{fig: Gamma F perturb} shows the features of this Liouvillian EP under such a perturbation.

We find that the Newton polygon, as shown in Figure~\ref{fig: Gamma F perturb} (a), exhibits two segments -- one with a slope $-1$ and another with a slope $-\frac{1}{3}$. The latter has a value $3$ as the projection on the horizontal axis. The shape of the Newton polygon is indicative of the Jordan canonical form, as we discussed above. Next, we look at the amoeba for the perturbations of $\gamma_f$. As presented in Figure~\ref{fig: Gamma F perturb}(b), we find that it shows two tentacles, one with slope 3 and the other with slope 1, both extending below. From these analyses, remarkably, we find that under a generic perturbation in $\gamma_f$, the four coalescing eigenvalues diverge into two groups -- one which varies linearly and the other three with a cube root structure. 

Our conclusions from the Newton polygon and tropical methods are further verified by direct numerical calculations. The numerical analysis on the perturbation of eigenvalues was done by direct diagonalization using Eigenvalues function implemented in Mathematica. The real and imaginary parts of the eigenvalues are plotted in Figure~\ref{fig: Gamma F perturb}(c) and (d), respectively, in the vicinity of the Liouvillian EP. The scaling, shown in Figure~\ref{fig: Gamma F perturb}(f), follows the expected cube-root dispersion. In Figure~\ref{fig: Gamma F perturb}(e), we present the cyclic permutation of the three eigenvalues with valuation $\frac{1}{3}$, which is an additional check on the exceptional behaviour. Thus, our approach faithfully captures the behavior of higher-order Liouvillian EPs.

\begin{figure}[!h]
   \fbox{\includegraphics[width=0.95\textwidth]{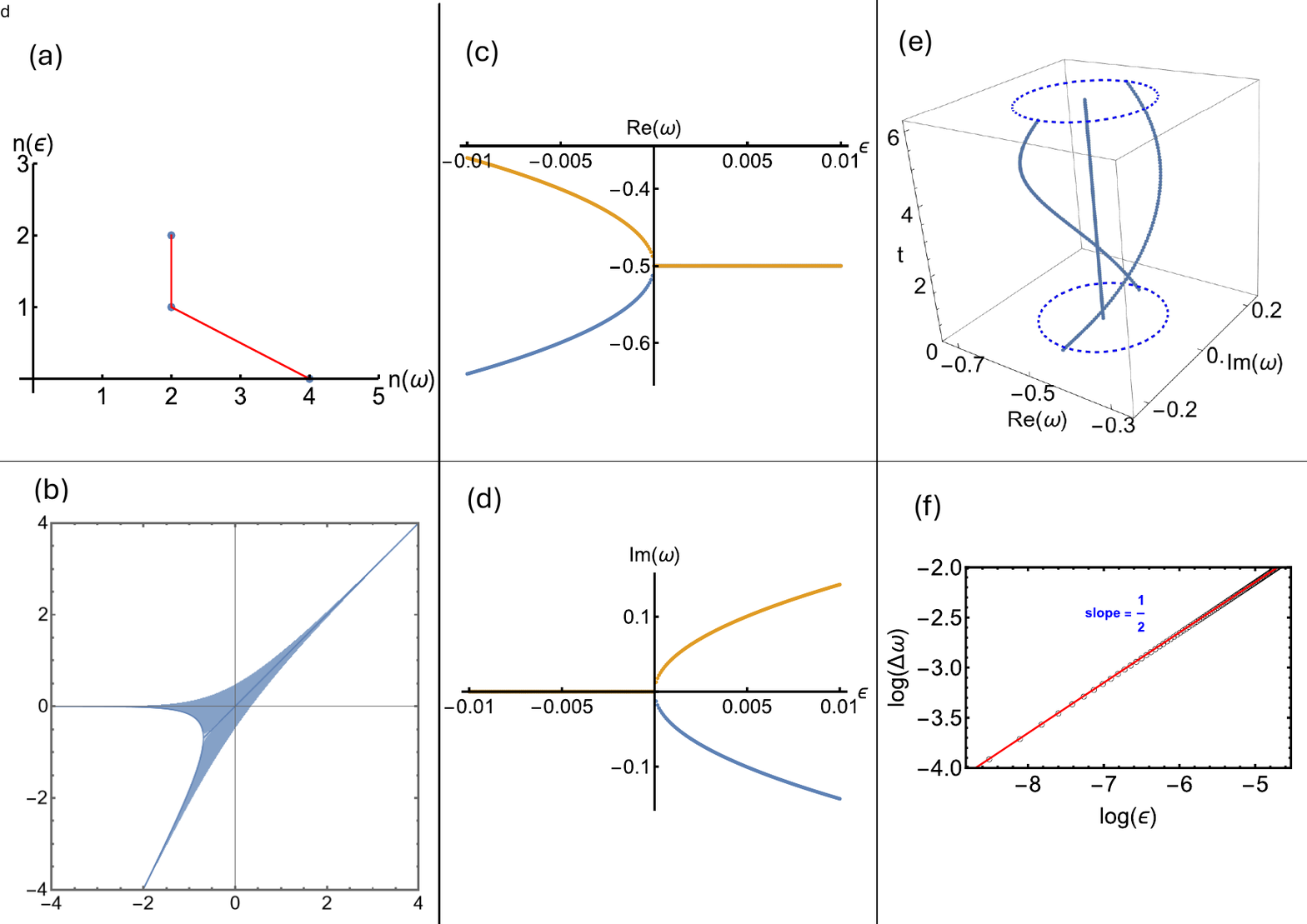} }
    \caption{\textbf{Hybrid Liouvillian exceptional points in dissipative qubit system with $J$ perturbation.} (a) The Newton polygon showing one segment with the slope of $-\frac{1}{2}$. The only other segment present is a vertical one. (b) The amoeba of the shifted characteristic polynomial at the Liouvillian EP. We see a tentacle branching downwards in the third quadrant with the slope of 2 and a purely horizontal tentacle branching to negative infinity along the horizontal axis. Amoebas therefore can predict invariant solutions as well, that is, zero roots. The (c) real and (d) imaginary part of the numerically evaluated energy eigenvalues near the Liouvillian EP showing two eigenvalues (green, orange) coalescing. Panel (e) shows the encircling of the Liouvillian EP, i.e., a plot of the eigenvalues of the matrix $\mathcal{L}_{eff}+0.01\mathcal{L}_J e^{it}$ with varying $t$. (f) Scaling plot of $\log(\omega-\omega_0)$ vs $\log (\epsilon)$, where the data points are taken from panels (c) and (d). A slope of $\frac{1}{2}$ indicates the square root scaling as predicted in (a) and (b). The parameters were chosen to be ($\gamma_f=0,\gamma_e=1,J=0.25$), and the perturbation matrix $\mathcal{L}_J$ is $\frac{\partial}{\partial J}\mathcal{L}_{eff}$ to consider small variations in $J$. Note that the behavior of the Liouvillian EP changes drastically by simply changing the perturbing parameter.} \label{fig: J Perturb}
\end{figure}

Next, let us consider the second case of the dissipative qubit model, where instead of $\gamma_f$, the value of the drive amplitude $J$ is perturbed. For this situation, our perturbation matrix can be constructed as $\frac{\partial }{\partial J} \mathcal{L}_{\text{eff}}$. We present the Newton polygon in Figure~\ref{fig: J Perturb}(a). Intriguingly, the Newton polygon shows a segment with slope $-\frac{1}{2}$ and no other segments with a finite slope. This means that under such a perturbation, out of the three coalesced eigenvalues, only two diverge with a square root topology. The other two eigenvalues, remain invariant under the perturbation, thus showing a valuation of $\infty$ or the slope of $-\infty$ in the Newton polygon. The amoeba, as constructed in Figure~\ref{fig: J Perturb}(b), also allows us to draw a similar conclusion. It has two tentacles diverging out in the negative infinity direction, one of which has a slope of $2$. This further ensures the square root topology. The other tentacle is horizontal, thus consolidating the invariance under the perturbation.

As before, we confirm our findings based on numerical computations. The real and imaginary parts of the eigenvalues near the Liouvillian EP are plotted in Figure~\ref{fig: J Perturb}(c) and (d), respectively. They follow the expected square root scaling (Figure~\ref{fig: J Perturb}(f)). The vorticity upon encircling the Liouvillian EP is shown in Figure~\ref{fig: J Perturb}(e). It clearly shows only two of the eigenvalues swapping. We note that here the straight vertical line is actually two such invariant lines which are coincident. However, they do not swap while encircling the Liouvillian EP. These numerical results further reaffirm our findings based on the Newton polygon and tropical methods.

Finally, we emphasize that the order of the same Liouvillian EP differs drastically with the chosen perturbation. This is elegantly captured by our methods of tropical analysis, as the choice of perturbation is central in our approach. In fact, this suggests that one can use Newton polygons and amoebas to obtain the necessary perturbation required to decrease the effective EP orders. Such perturbations are called non-generic~\cite{kato2013perturbation}, and, in conjunction with our methods, could be useful to design Liouvillian and Hamiltonian EPs of the desired order.

\section{Summary and Outlook}\label{sec:summary}

In this paper, we have shown that our methods of Newton polygons and tropical geometry can directly characterize Liouvillian EPs. Using the dissipative spin$-1/2$ model, we demonstrated how our approach allows us to obtain the exhaustive set of Liouvillian EPs, as well as fine tune and characterize them by analyzing the Newton polygons and amoebas. Furthermore, using the dissipative superconducting qubit model as an example, we showed that our approach captures the anisotropy and order of the Liouvillian EPs depending on the form of the perturbation. We supplemented our analytical analyses with numerical calculations of scaling of the eigenvalues as well as their exchange upon encircling around the Liouvillian EPs. In future, exploring non-linear~\cite{bai2023nonlinear} and non-Markovian~\cite{lin2025non} variants of EPs with our methods could be fruitful directions. We are hopeful that our analytical approach will be a useful tool to aid in the design and discovery of Liouvillian EPs.

\section{Appendix}\label{sec:appendix}

\subsection*{Obtaining the amoeba of a polynomial}
We summarize the main steps in obtaining the amoeba for the equation in Fig 1. We reiterate that the amoeba is the image of the set of solutions under the logarithmic map. We start with\\

\begin{equation}
    f(\omega,\epsilon) = -\epsilon + 9\omega^2 -  \omega^4 = 0,
\end{equation}
which can be exactly solved in $\epsilon$. Take the absolute logarithm, to obtain
\begin{equation}
\log{|\epsilon|} = \log |9 \omega^2 - \omega^4|.
\end{equation}
We now write $\omega$ in its polar form $\omega=|\omega| \exp{i \theta}$. Since the horizontal and vertical axes in the amoeba plot denote $\log|\omega| = x$ and $\log|\epsilon| = y$, respectively, we have
\begin{align}
y &= \log |9 \exp({x+i\theta})^2 - \exp({x+i\theta})^4|, \nonumber \\
 &= 2x + \log|9 - \exp(2x + 2i\theta)|.
\end{align}
Here $\theta$ takes value from $0$ to $2 \pi$. So the grey area is parametrized by two variables $x$ and $\theta$, such that
\begin{equation}
(x,y) = (x,2x + \log|9 - \exp(2x + 2i\theta)|).
\end{equation}
Since $\log|x|$ goes towards infinity in cases of either $x$ being close to zero or $x$ being close to infinity, the asymptotes of the logarithm map denotes the three asymptotic solutions of $f$. First is the large $\omega$ and large $\epsilon$ regime, giving
\begin{equation}
    y \approx 2x + \log|e^{2x} e^{2 i\theta}| = 4x,
\end{equation}
or,
\begin{equation}
    \epsilon \sim \omega^4.
\end{equation}
Then we have the small $\omega$ and small $\epsilon$ regimes, where the quadratic term dominates in Eq. (2), giving
\begin{equation}
     y \approx 2x + \log|9|,
\end{equation} 
or,
\begin{equation}
    \epsilon \sim 9\omega^2.
\end{equation}
There is a third case when $\omega \rightarrow 3, \epsilon \rightarrow0$, a finite $\omega$ small $\epsilon$ regime. Here as $x = \log|\omega|\rightarrow\log|3|$,
\begin{align}
y &= 2\log(3) + \log|9 - 9(e^{2i\theta})|, \nonumber \\
 &= 4\log(3) + \log|1-e^{2i\theta}|, \nonumber \\
 &= 4\log(3) + \log2+ \log|\sin{\theta}|.
\end{align}
Since $\lim_{\theta \rightarrow0}\log|\sin{\theta}| \rightarrow -\infty$, we get a vertical asymptotic line that extends towards $-\infty$. This is the third asymptote. We see that the amoeba helps identify the asymptotic regimes of the equation. In our case, since we are concerned with small perturbations, we only need to consider the asymptotes that go towards the $-\infty$ along $x$-axis.

\subsection*{Characteristic polynomial of spin$-1/2$ dissipative system}

In this appendix, we provide the lengthy expressions for the coefficients in the characteristic polynomial of the spin$-1/2$ dissipative system. The coefficient of $\omega$ is given by

\begin{eqnarray}
\begin{aligned}
&0.25 \, \gamma_{-}^3 
+ 1.5 \, \gamma_{-}^2 \gamma_{x} 
+ 2 \, \gamma_{-} \gamma_{x}^2 
+ 1.5 \, \gamma_{-}^2 \gamma_{y} 
+ 8 \, \gamma_{-} \gamma_{x} \gamma_{y} 
+ 8 \, \gamma_{x}^2 \gamma_{y} 
 + 2 \, \gamma_{-} \gamma_{y}^2 
+ 8 \, \gamma_{x} \gamma_{y}^2 
+ 1 \, \gamma_{-} \Omega^2 
+ 2 \, \gamma_{x} \Omega^2 
+ 2 \, \gamma_{y} \Omega^2 \nonumber \\
&\quad + 2.5 \, \gamma_{-}^2 \omega_{0} 
+ 10 \, \gamma_{-} \gamma_{x} \omega_{0} 
+ 8 \, \gamma_{x}^2 \omega_{0} 
+ 10 \, \gamma_{-} \gamma_{y} \omega_{0} 
+ 24 \, \gamma_{x} \gamma_{y} \omega_{0} 
+ 8 \, \gamma_{y}^2 \omega_{0} \nonumber \\
&\quad + 2 \, \Omega^2 \omega_{0} 
+ 6 \, \gamma_{-} \omega_{0}^2 
+ 12 \, \gamma_{x} \omega_{0}^2 
+ 12 \, \gamma_{y} \omega_{0}^2 
+ 4 \, \omega_{0}^3.
\end{aligned}
\end{eqnarray}

The coefficient of the constant term is obtained to be 

\begin{eqnarray}
\begin{aligned}
&0.25 \, \gamma_{-}^3 \omega_{0} 
+ 1.5 \, \gamma_{-}^2 \gamma_{x} \omega_{0} 
+ 2 \, \gamma_{-} \gamma_{x}^2 \omega_{0} 
+ 1.5 \, \gamma_{-}^2 \gamma_{y} \omega_{0} 
+ 8 \, \gamma_{-} \gamma_{x} \gamma_{y} \omega_{0} 
+ 8 \, \gamma_{x}^2 \gamma_{y} \omega_{0} \nonumber \\
&\quad + 2 \, \gamma_{-} \gamma_{y}^2 \omega_{0} 
+ 8 \, \gamma_{x} \gamma_{y}^2 \omega_{0} 
+ 1 \, \gamma_{-} \Omega^2 \omega_{0} 
+ 2 \, \gamma_{x} \Omega^2 \omega_{0} 
+ 2 \, \gamma_{y} \Omega^2 \omega_{0} \nonumber \\
&\quad + 1.25 \, \gamma_{-}^2 \omega_{0}^2 
+ 5 \, \gamma_{-} \gamma_{x} \omega_{0}^2 
+ 4 \, \gamma_{x}^2 \omega_{0}^2 
+ 5 \, \gamma_{-} \gamma_{y} \omega_{0}^2 
+ 12 \, \gamma_{x} \gamma_{y} \omega_{0}^2 
+ 4 \, \gamma_{y}^2 \omega_{0}^2 \nonumber \\
&\quad + 1 \, \Omega^2 \omega_{0}^2 
+ 2 \, \gamma_{-} \omega_{0}^3 
+ 4 \, \gamma_{x} \omega_{0}^3 
+ 4 \, \gamma_{y} \omega_{0}^3 
+ \omega_{0}^4.
\end{aligned}
\end{eqnarray}

Equating these coefficients to zero allows us to eliminate these in the Newton polygon and scan for the second-order Liouvillian EPs. First we eliminate $\omega_0$ by taking the resultant, and then solve the resultant for one of the variables, say $\gamma_x$. We obtained four solution regimes as 

\begin{equation}
    \{\gamma_x \to \tfrac{1}{2}(-\gamma_- - 2\gamma_y)\}, \;
\{\gamma_x \to \gamma_y \pm \Omega\}, \;
\{\gamma_x \to \tfrac{1}{4}\tfrac{-\gamma_-^2 - 4\gamma_- \gamma_y - 4\Omega^2}{\gamma_- + 4\gamma_y}\}.
\end{equation}

For each of them, we substitute for $\gamma_x$ into the respective characteristic polynomial and construct the Newton polygon. Using this approach, we find that the fourth region is not a Liouvillian EP but rather a diabolic point in the spectrum. 

By performing a similar procedure with a different variable, we can scan for regions independent of $\gamma_x$. By repeating this procedure no new solutions were obtained.

\subsection*{Comparison with the Jordan canonical decomposition}

Jordan canonical decompositions provide a clear and unique classification tool for matrices based on their geometric and algebraic multiplicities. Every square matrix is similar under a transformation to a block diagonal matrix called the Jordan canonical form. These blocks are called Jordan blocks and denote invariant subspaces of the matrix. 

There is a one-to-one correspondence between the Jordan canonical form and the structure of the Newton-Puiseux diagram upon a {\it generic} perturbation. 

Consider an $N \times N$ matrix, $\mathcal{L}_0$, with complex entries, which in its Jordan canonical form, decomposes into $k$ Jordan blocks with of size $s_i$ and eigenvalue $\lambda_i$, $i=1,2,3\cdots k \text{ with } \sum_{i=1}^k s_i=N$. Under a generic perturbation, each Jordan block of size $s_i$ gives rise to $s_i$ eigenvalues admitting Puiseux expansions $\lambda-\lambda_i \sim \epsilon^{1/s_i}$, forming a cyclic splitting in the complex plane. This is a standard result in the Lidskii-Vishik-Lyusternik perturbation theory~\cite{lidskii1966perturbation,a3referee}

This is exactly the same information that is provided by the Newton polygon. If we construct the Newton polygon of $\det(\mathcal{L}_0 +\epsilon \mathcal{L}_1- I (\omega -\lambda))$, we can find all Jordan blocks with diagonal entry $\lambda$.

For the rest of the section, we work within an eigenspace, that is, assume $\mathcal{L}_0$ has all the same eigenvalue $\lambda$. Then, we can obtain the Jordan canonical form of $H_0$ from the Newton-Puisuex diagram using the following procedure:
\begin{itemize}
    \item Consider all segments of the Newton polygon diagram that subtends a projection of 1 unit on the vertical axis. Here, by segment, we mean a line connecting two points in the Newton polygon plot that is to the left and below all other points.\
    \item The number of such segments is the number of Jordan blocks with diagonal entry $\lambda$. 
    \item The projection on the horizontal axis will denote the size of the Jordan block.
\end{itemize}

\begin{figure}
    \centering
    \fbox{\includegraphics[width=0.85\linewidth]{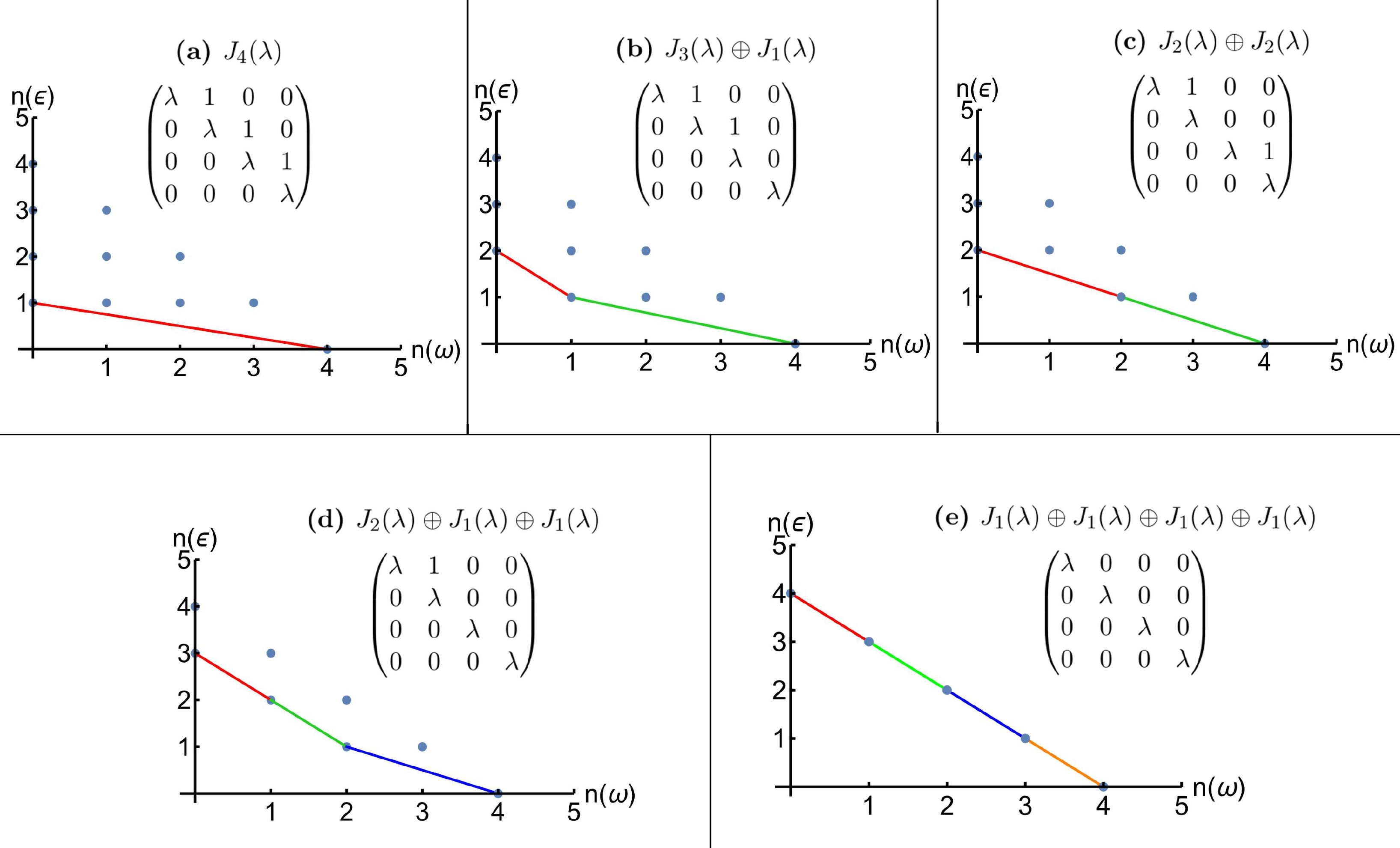}}
    \caption{\textbf{Newton polygons and Jordan canonical forms.} The isomorphism dictionary between Newton polygons of matrices under a generic perturbation and their Jordan canonical forms for $N=4$ matrices. Each of the colored segments indicates a different Jordan block. Out of all these cases, only panel (e) shows a diabolic point; the rest are all exceptional points.}
    \label{fig:jordannewton}
\end{figure}

Figure~\ref{fig:jordannewton} illustrates this isomorphism for the case of $N=4$. Note here that the Newton polygon of the polynomial $\det(\mathcal{L}_0+\epsilon \mathcal{L}_1 - I( \omega-\lambda))$ is plotted next to the Jordan form of $\mathcal{L}_0$. 

As can be seen from Figure~\ref{fig:jordannewton}, our model II shows a $J_1 \oplus J_3$ type Hamiltonian, with perturbing $\mu$ and $J$ resulting into generic and non-generic perturbations, respectively. Model I on the other hand describes a $J_2$ type EP. 

We see that Newton polygons provide an analytic method to perform the Jordan decomposition. Amoebas, being dual to Newton polygons, can also be used and provide a purely numerical approach to perform Jordan decompositions. As the matrix size gets large amoebas potentially provide a complementary approach for numerical studies, similar to random matrix techniques.

A non-generic perturbation can thus be defined as a perturbation matrix $\mathcal{L}_1$ that changes the expected structure of the Newton polygon as compared to its Jordan form. It can do this by making one of vertices of the segments vanish. If these coefficients are given by $g_1(\mathcal{L}_1),g_2(\mathcal{L}_1) \cdots g_k (\mathcal{L}_1)$, then the set of all non-generic matrices will then be given by the union of their roots.

Furthermore, by equating the coefficients of vertices to zero, we can also obtain the formulae for a general region of non-generic perturbation. This can assist in obtaining a non-generic perturbation matrix for a different desired eigenvalue unfolding in an experimental setting.

\subsection*{Obtaining complete eigenvalue unfolding structure}

Newton polygons and amoeba techniques are methods that not only solve for the simplest Puiseux behavior but also the entire solution in terms of an infinite series~\cite{howtocomputepuiseux}. Given an algebraic polynomial of the form $f(\omega,\epsilon)$, one can solve for $\omega$ as a Puiseux series in $\epsilon$ as 
$$\omega=\omega_0 + c_1 \epsilon^{\gamma_1} + c_2\epsilon^{\gamma_1+\gamma_2} +\cdots .$$ The procedure for doing this is as follows
\begin{itemize}
    \item First use the aforementioned methods to obtain the valuation $\gamma_1$ in the solution.
    \item Obtain $f(\omega_0+c_1 \epsilon^{\gamma_1},\epsilon)$, and equate the coefficient of the lowest order term of $\epsilon$ to zero. This will give a polynomial equation for $c_1$ of degree $d$, where $d$, as expected from the Newton polygon theorems, is the projection of the polygon segment on the horizontal axis.
    \item Then construct the polynomial $f_1(\omega',\epsilon) = \epsilon^{-\beta}f(w_0+c_1 (\epsilon^{\gamma_1} + \omega'),\epsilon)$ where $\beta$ is the vertical axis intercept of the corresponding segment.
    \item Reiterate the process to find the leading coefficient of $\omega_1$ and so forth continuing to generate a series.
\end{itemize}

As an illustration, let us obtain the concrete ring unfolding of the three-ring in model II under the $\gamma_f$ perturbation (Figure 2).
Here, $\gamma_1$ is the valuation $\frac{1}{3}$, and $\omega_0=-\frac{1}{2}$. Obtaining $f(\omega_0+c_1 \epsilon^{\gamma},\epsilon)$ and collecting the lowest order terms of $\epsilon$, we find
\begin{equation}
    f(\omega_0+c_1 \epsilon^{\gamma},\epsilon) = (-\frac{1}{8} + c_1^3) \epsilon^{4/3} + \mathcal{O}(\epsilon^{5/3} ).
\end{equation}
Equating the coefficient to zero, we obtain $ c_1= \frac{1}{2} \times (1, e^{\frac{2\pi}{3}},e^{\frac{4\pi}{3}})$ corresponding to the three coalescing eigenvalues. We see that one of them remains purely real as was seen in Figure 2.
We illustrate the second iteration for $\omega = -\frac{1}{2} + \frac{1}{2} \epsilon^{1/3} + \omega_1$. 
\begin{equation}
\begin{aligned}
f_1(\omega',\epsilon)
&= \epsilon^{-\beta}
   f\!\left(
     \omega_0 + c_1\bigl(\epsilon^{\gamma_1} + \omega'\bigr),
     \epsilon
   \right) \\[0.5em]
&= \omega'
 + \omega'^3 \left(2 \epsilon^{2/3} + 2\right) \\[0.3em]
&\quad
 + \omega'^2 \left(
     \frac{5}{4}\epsilon^{4/3}
     + 3\epsilon^{2/3}
     + \frac{1}{2}\epsilon^{1/3}
     + \frac{3}{2}
   \right) \\[0.3em]
&\quad
 + \omega' \left(
     \frac{5}{4}\epsilon^{4/3}
     + \frac{3}{2}\epsilon^{2/3}
     + \frac{1}{4}\epsilon^{2}
     + \frac{1}{2}\epsilon
     + \frac{1}{2}\epsilon^{1/3}
     + \frac{3}{8}
   \right) \\[0.3em]
&\quad
 + \frac{1}{8}\epsilon^{5/3}
 + \frac{5}{16}\epsilon^{4/3}
 + \frac{3}{16}\epsilon^{2/3}
 + \frac{1}{8}\epsilon^{2}
 + \frac{1}{4}\epsilon
 + \frac{1}{8}\epsilon^{1/3}.
\end{aligned}
\end{equation}

The Newton polygon when constructed shows a single segment with a slope of $-\frac{1}{3}$ and horizontal axis projection of 1 unit. Calculating $f_1(c_2 \epsilon^{1/3},\epsilon)$ yields,
\begin{align}    
f_1(c_2 \epsilon^{1/3},\epsilon)&=\frac{1+3c_2}{8} \epsilon^{1/3} + \mathcal{O}(\epsilon^{2/3}),\\
\implies c_2&= -\frac{1}{3}.
\end{align}
One can repeat this procedure for the other two complex eigenvalues as well. However, we know that these eigenvalues perform a cyclic permutation under encircling of the EP. Thus, the coefficient of $\epsilon^p$ for the complex eigenvalues is simply obtained as $c_p\, e^\frac{i 2 p \pi}{3}$ and $c_p\, e^\frac{i 4 p \pi}{3}$, where $c_p$ is the corresponding coefficient of the first eigenvalue.

Summarizing the results up to the second term in $\epsilon$, we get
\begin{align*}
    \omega^1=& -\frac{1}{2} + \frac{1}{2} \epsilon^{1/3}- \frac{1}{3} \, \epsilon^{2/3} + \cdots\\
    \omega^2=& -\frac{1}{2} + \frac{e^{\frac{i 2 \pi}{3}}}{2} \epsilon^{1/3}- \frac{e^{\frac{i 4 \pi}{3}}}{3} \, \epsilon^{2/3} + \cdots\\
    \omega^3=& -\frac{1}{2} + \frac{e^{\frac{i 4 \pi}{3}}}{2} \epsilon^{1/3}-\frac{e^{\frac{i 8 \pi}{3}}}{3} \, \epsilon^{2/3} + \cdots
\end{align*}

with the superscript denoting the eigenvalue index.

\section*{Acknowledgments} 

P.S. thanks Innovation in Science Pursuit for Inspired Research (INSPIRE) scholarship. A.N. acknowledges support from the DST MATRICS grant (MTR/2023/000021). We thank R. Sarkar for useful discussions. A.N. thanks A. Banerjee, R. Jaiswal and M. Manjunath for related collaborations.

\bibliography{bibliography}
\end{document}